\RequirePackage{fix-cm}
\documentclass[twocolumn]{svjour3}          
\smartqed  
\usepackage{graphicx}
\usepackage{color}
\usepackage{color}
\begin{document}

\title{Spatial correlation of fluctuations in multi-component superconducting systems}



\author{Teet \"{O}rd \and Artjom Vargunin \and K\"{u}llike R\"{a}go 
}


\institute{
Teet \"{O}rd, K\"{u}llike R\"{a}go,  Artjom Vargunin \at
            Institute of Physics, University of Tartu \\ T\"{a}he 4, 51010 Tartu, Estonia \\
               \email{teet.ord@ut.ee}          
}

\date{Received: date / Accepted: date}

\maketitle

\begin{abstract}
We show that spatial variation and correlation of superconductivity fluctuations in a two-band model are scaled by two characteristic lengths. This results in substantially more complicated picture compared to one-band systems. In particular, short-range correlations are always present in a two-band scenario, even near the phase transition point.
\end{abstract}

\section{Introduction}

The multi-band superconductivity mechanisms with interband pairing channels have been became more and more actual \cite{bianconi2,aoki,bianconi3}. The rich physics included here is reflected by the unusual properties of relevant compounds.
In particular, the discovery of novel vortex structure in magnesium diboride \cite{moshchalkov}, interpreted as type-1.5 superconductivity, points unambiguously to the non-trivial modification of coherency in multi-gap systems \cite{babaev1}. Recently the existence of type-1.5 superconductivity was suggested also in strontium ruthenate \cite{babaev4}.

The various aspects of spatially inhomogeneous multi-gap superconducting ordering were considered in Refs. \cite{babaev1,babaev,babaev2,shanenko1,babaev3,ord1,litak2012,shanenko2,litak2012a,varg}. In this paper we examine the correlations and spatial variation of fluctuations in a two-band superconductor with intraband and interband pairings.

\section{Correlation lengths in two-band superconductivity}

The spatially inhomogeneous two-band superconductivity is described (see e. g. \cite{ord1}) by the following gap equations in the vicinity of critical point
\begin{eqnarray}\label{eq1}
\Delta_{\alpha }(\mathbf{r})=&-&\sum_{\alpha '}W_{\alpha \alpha'}\rho_{\alpha '}\biggl[g(T)-\nu
\left|\Delta_{\alpha '}(\mathbf{r})\right|^{2} \nonumber \\
&+&\beta_{\alpha '}\nabla^{2}\biggr]\Delta_{\alpha '}(\mathbf{r}) \, .
\end{eqnarray}
Here
\begin{eqnarray}\label{eq2}
g(T)=\ln\left(\frac{1.13\hbar\omega_{D}}{k_{B}T}\right) \, ,
\end{eqnarray}
\begin{eqnarray}\label{eq3}
\beta_{\alpha }=\frac{7\zeta(3)\hbar^{2}v^{2}_{F\alpha}}{48(\pi k_{B}T_{c})^{2}} \, ,
\end{eqnarray}
\begin{eqnarray}\label{eq4}
\nu=\frac{7\zeta(3)}{8(\pi k_{B}T_{c})^{2}} \, ,
\end{eqnarray}
$v_{F\alpha}$ are the Fermi velocities and $\rho_{\alpha}$ are the electron densities of states at the Fermi level in the corresponding bands, $\alpha=1,2$.
It is supposed that the superconducting ordering is created by the intraband effective electron-electron attractions with constants $W_{11,22}<0$ and by interband pair-transfer interaction (constant $W_{12}=W_{21}$). These interaction channels are supposed to be operative in the energy layer $\pm \hbar\omega_{D}$ around the Fermi level intersecting both bands. The temperature of the bulk superconducting phase transition $T_{c}$ follows from the linearized equations (\ref{eq1}) in the spatially homogeneous case.

We consider the small deviation from the bulk superconductivity representing $\Delta_\alpha(\mathbf{r})=\Delta^\infty_\alpha+\left|\Delta^{\infty}_{\alpha}\right|f_{\alpha}(\mathbf{r})$ where $\Delta^{\infty}_{\alpha}$ are the bulk gaps. The system of linearized equations for the fluctuations $f_{\alpha}(\mathbf{r})$ follows from Eq. (\ref{eq1}), namely
\begin{eqnarray}\label{eq5}
&&\xi^{2}_{1}\nabla^{2}f_{1}(\mathbf{r})=f_{1}(\mathbf{r})+\Xi_{12}f_{2}(\mathbf{r}), \nonumber\\
&&\xi^{2}_{2}\nabla^{2}f_{2}(\mathbf{r})=f_{2}(\mathbf{r})+\Xi_{21}f_{1}(\mathbf{r}),
\end{eqnarray}
with
\begin{eqnarray}\label{eq6}
\xi_{\alpha}=\sqrt{\frac{K_{\alpha}}{A_{\alpha}}} \, , \,\,\,\,\,  \Xi_{12,21}=\frac{c}{A_{1,2}}\frac{\left|\Delta^{\infty}_{2,1}\right|}{\left|\Delta^{\infty}_{1,2}\right|} \, ,
\end{eqnarray}
\begin{eqnarray}\label{eq7}
K_{\alpha}=\rho_{\alpha}\beta_{\alpha} \, , \,\,\,\,\, A_{\alpha}=a_{\alpha}+3b_{\alpha}|\Delta_{\alpha}^{\infty}|^{2}  \, ,
\end{eqnarray}
\begin{eqnarray}\label{eq8}
c=\frac{W_{12}}{W_{12}^{2}-W_{11}W_{22}} \, ,
\end{eqnarray}
\begin{eqnarray}\label{eq9}
a_{1,2}(T)=\frac{W_{22,11}}{W_{12}^{2}-W_{11}W_{22}}-\rho_{1,2}g(T) \, , \,\,\,\,\, b_{\alpha}=\rho_{\alpha}\nu \, .
\end{eqnarray}
If interband interaction channel is absent, $\Xi_{12}= \Xi_{21}=0$ and the system (\ref{eq5}) splits into two independent equations. For this case $\xi_{1,2}$ are the coherence lengths which describe the spatial behaviour of superconductivity in the autonomous bands.

For finite interband coupling the quantities $\xi_{1,2}$ cannot be treated as relevant coherence lengths. Instead, one obtains \cite{ord1} on the basis of Eqs. (\ref{eq5}) the length scales $\xi_{\pm}$ for the spatial variation of $f_{\alpha}(\mathbf{r})$, and
\begin{eqnarray}\label{eq10}
\xi_{\pm}^{-2}&=&\frac{1}{2}\Bigg\{\xi_{1}^{-2}+\xi_{2}^{-2} \nonumber \\
&\pm&\sqrt{\left(\xi_{1}^{-2}-\xi_{2}^{-2}\right)^{2}+4\frac{c^{2}}{A_{1}A_{2}}\xi_{1}^{-2}\xi_{2}^{-2}}\Bigg\} \, .
\end{eqnarray}
The coherence lengths found cannot be attributed to the separate bands. Their temperature dependencies are shown in Fig. 1. As $\frac{c^{2}}{A_{1}A_{2}}=1$ at the phase transition temperature $T_{c}$, one observes the critical (diverging) behaviour of $\xi_{-}(T)$ near $T_{c}$, while $\xi_{+}(T)$ remains finite.
\begin{figure}[!b]
\begin{center}
\resizebox{0.8\columnwidth}{!} {\includegraphics[angle=-90]{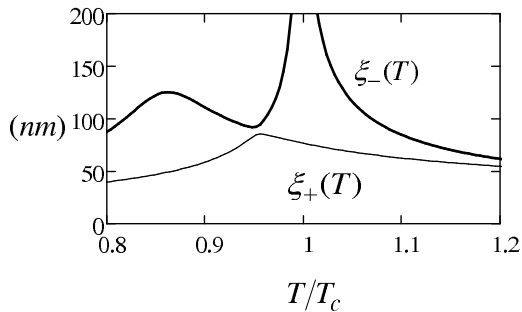}}
\caption{The coherence lengths $\xi_{-}$ and $\xi_{+}$ \textit{vs} temperature. Parameters: $\rho_{1}W_{11}=-0.18 $, $\rho_{2}W_{22}=-0.174$, $|W_{12}|\sqrt{\rho_{2}\rho_{1}}=0.0003$, $\hbar\omega_{D}=0.3 \, eV$, $v_{F1}=4\times 10^{5} \, m/s$, $v_{F2}=5\times 10^{5} \, m/s$. } \label{f1}
\end{center}
\end{figure}
The maximum of $\xi_{-}(T)$ below $T_{c}$ is characteristic feature of a superconductor with tiny interband pair-transfer interaction. That non-monotonicity is correlated with the temperature where the band with weaker superconductivity becomes active. The second peak of $\xi_{-}(T)$ disappears with increase of interband coupling.

\section{Spatial behaviour of fluctuations}

By considering the dependency on $x$-coordinate only, the general solution of the system (\ref{eq5}) has  the form
\begin{eqnarray} \label{eq11}
f_{1}(x)&=&p_{1}^{-}e^{-x/\xi_{-}}+p_{1}^{+}e^{-x/\xi_{+}} \, , \nonumber \\
f_{2}(x)&=&p_{2}^{-}e^{-x/\xi_{-}}+p_{2}^{+}e^{-x/\xi_{+}} \, ,
\end{eqnarray}
where
\begin{eqnarray}\label{eq12}
p^{-}_{1}&=&\frac{b_{+}f_{1}(0)-f_{2}(0)}{b_{+}-b_{-}}=\frac{p^{-}_{2}}{b_{-}} \, , \nonumber \\
p^{+}_{1}&=&\frac{f_{2}(0)-b_{-}f_{1}(0)}{b_{+}-b_{-}}=\frac{p^{+}_{2}}{b_{+}}
\end{eqnarray}
with
\begin{eqnarray}\label{eq13}
b_{\pm}=\frac{\xi^{2}_{1}\xi^{-2}_{\pm}-1}{\Xi_{12}}=\frac{\Xi_{21}}{\xi^{2}_{2}\xi^{-2}_{\pm}-1} \, .
\end{eqnarray}
The temperature dependencies of the amplitudes $p_{1,2}^{\pm}$ are depicted in Fig. 2 for weak interband pairing.  For $T\rightarrow T_{c}$ the spatial behaviour in the band with stronger superconductivity is scaled predominantly by the critical coherence length ($p_{1}^{-} \gg p_{1}^{+}$). In this case the superconducting ordering is very close to the one-band situation  where only one (critical) coherence length is of relevance.
\begin{figure}[!b]
\begin{center}
\resizebox{0.7\columnwidth}{!} {\includegraphics[angle=-90]{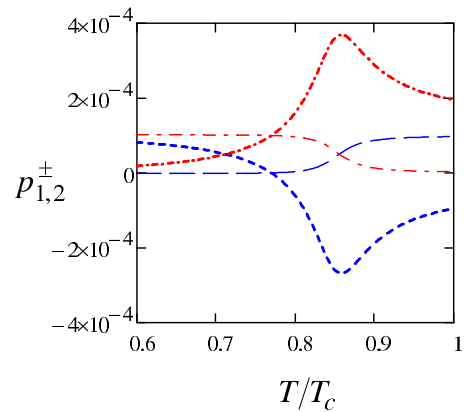}}
\caption{The coefficients $p_{1}^{-}$ (thin dashed blue line), $p_{1}^{+}$ (thin dashed-dotted red line), $p_{2}^{-}$ (thick dashed blue line) and $p_{2}^{+}$ (thick dashed-dotted red line) \textit{vs} temperature for $|W_{12}|\sqrt{\rho_{1}\rho_{2}}=0.003$. The other parameters: $f_{1}(0)=f_{2}(0)=10^{-4}$, $\rho_{1}W_{11}=-0.23$, $\rho_{2}W_{22}=-0.2$, ,  $\hbar\omega_{D}=0.3 \, eV$, $v_{F1}=4\times 10^{5} \, m/s$, $v_{F2}=5\times 10^{5} \, m/s$. } \label{f2}
\end{center}
\end{figure}
\begin{figure}[!t]
\begin{center}
\resizebox{0.7\columnwidth}{!} {\includegraphics[angle=-90]{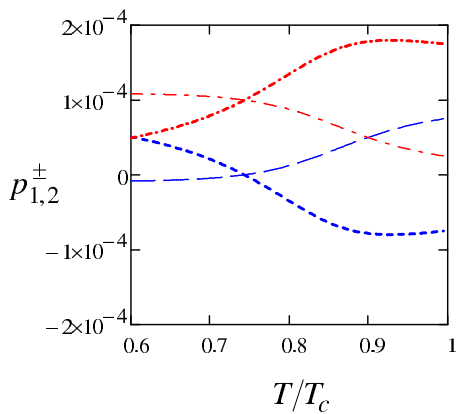}}
\caption{The coefficients $p_{1}^{-}$ (thin dashed blue line), $p_{1}^{+}$ (thin dashed-dotted red line), $p_{2}^{-}$ (thick dashed blue line) and $p_{2}^{+}$ (thick dashed-dotted red line) \textit{vs} temperature for $|W_{12}|\sqrt{\rho_{1}\rho_{2}}=0.01$.  The other parameters as in Fig. \ref{f2}. } \label{f3}
\end{center}
\end{figure}
Larger discrepancy between intrinsic intraband critical points and the weakness of interband interaction play in favour of that trend. With the lowering of the temperature (Fig. \ref{f2}) or with increase of interband pairing (Fig. \ref{f3}) the non-critical channel becomes important for the band considered.

\begin{figure}[!h]
\begin{center}
\resizebox{0.7\columnwidth}{!} {\includegraphics[angle=-90]{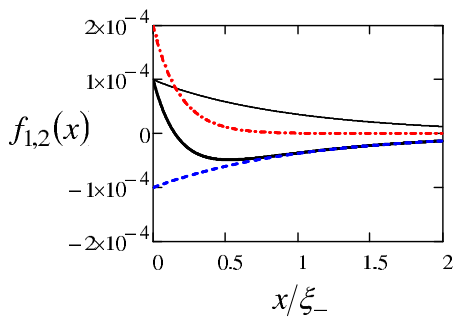}}
\caption{The dependence of $f_{1}(x)$ (thin black line) and $f_{2}(x)$ (thick black line) together with critical (thick dashed blue line) and non-critical (thick dashed-dotted red line) contributions to $f_{2}(x)$ for $T=0.99T_{c}$ and $|W_{12}|\sqrt{\rho_{1}\rho_{2}}=0.003$. The other parameters as in Fig. \ref{f2}. } \label{f4}
\end{center}
\end{figure}

For the band with weaker superconductivity both critical and non-critical channels contribute near $T_{c}$ ($p_{2}^{+}\sim -p_{2}^{-}$). From Fig. \ref{f4} we clearly see that for shorter distances $f_{2}(x)$ scales by $\xi_{+}$, while for longer distances by $\xi_{-}$. Moreover, the interplay of critical and non-critical contributions can result in the non-monotonic dependence of $f_{2}(x)$.

\section{Correlation functions of two-band superconductivity}

Critical and non-critical coherence lengths (\ref{eq10}) appear as scaling factors in the expressions for the correlation functions of superconductivity fluctuations
\begin{eqnarray}\label{eq14}
\Gamma_{\alpha\alpha'}\left(\mathbf{r},\mathbf{r}'\right)=\left\langle \eta_{\alpha}\left(\mathbf{r}\right)\eta^{\ast}_{\alpha'}\left(\mathbf{r}'\right)\right\rangle
- \left\langle \eta_{\alpha}\left(\mathbf{r}\right)\right\rangle \left\langle\eta^{\ast}_{\alpha'}\left(\mathbf{r}'\right)\right\rangle \, ,
\end{eqnarray}
where $\eta_{\alpha}\left(\mathbf{r}\right)=\delta_{\alpha}\left(\mathbf{r}\right)-\Delta_{\alpha}^{\infty}$ and $\delta_{\alpha}\left(\mathbf{r}\right)$ are the non-equilibrium gap order parameters. By using the statistics, determined by the non-equilibrium free energy, one finds in the Gaussian approximation \cite{varg}
\begin{eqnarray}\label{eq15}
\Gamma_{\alpha\alpha^\prime}(\mathbf{r-r^\prime})=\Gamma_{\alpha\alpha^\prime}^+(\mathbf{r-r^\prime})+\Gamma_{\alpha\alpha^\prime}^-(\mathbf{r-r^\prime})
\end{eqnarray}
with
\begin{eqnarray}\label{eq16}
\Gamma_{11}^\pm(\mathbf{r-r^\prime})&=&\mp\frac{k_\mathrm{B}T}{8\pi K_\alpha}\frac{\xi_\mp^2(\xi_\pm^2-\xi_{2}^2)}{\xi_{2}^2(\xi_-^2-\xi_+^2)}\frac{\exp\left(-\frac{|\mathbf{r-r^\prime}|}{\xi_\pm}\right)}{|\mathbf{r-r^\prime}|} \, , \nonumber \\
\Gamma_{22}^\pm(\mathbf{r-r^\prime})&=&\mp\frac{k_\mathrm{B}T}{8\pi K_\alpha}\frac{\xi_\mp^2(\xi_\pm^2-\xi_{1}^2)}{\xi_{1}^2(\xi_-^2-\xi_+^2)}\frac{\exp\left(-\frac{|\mathbf{r-r^\prime}|}{\xi_\pm}\right)}{|\mathbf{r-r^\prime}|} \, , \nonumber \\
\Gamma_{12}^\pm(\mathbf{r-r^\prime})&=&\mp\frac{k_\mathrm{B}T}{8\pi K_1K_2}\frac{\xi_+^2\xi_-^2c}{\xi_-^2-\xi_+^2}\frac{\exp\left(-\frac{|\mathbf{r-r^\prime}|}{\xi_\pm}\right)}{|\mathbf{r-r^\prime}|} \, .
\end{eqnarray}

As interband pairing vanishes, $\Gamma_{\alpha\alpha}$ transforms into the correlation function of independent band, scaled by the relevant coherence length, and $\Gamma_{12}\rightarrow0$. If interband interaction is present, the picture becomes more complicated. For shorter distances $|\mathbf{r-r^\prime}|\ll\xi_+<\xi_-$ the functions $\Gamma_{\alpha\alpha}$ decrease in space as $|\mathbf{r-r^\prime}|^{-1}$, and this dependence may be caused  by the different contributions in the bands stemming from critical or non-critical coherency channels. As the distance separating fluctuations increases, long range correlations, determined by $\xi_-$, become dominating. Therefore, there exists a possibility for the interchange of the leading role between the critical and non-critical contributions in the spatial behaviour of $\Gamma_{\alpha\alpha}$. At $T\approx T_{c}$ the short-range correlations $\sim \left|\mathbf{r}-\mathbf{r}'\right|^{-1}\exp\left(-\frac{\left|\mathbf{r}-\mathbf{r}'\right|}{\xi_{+}}\right)$ are also involved. In the one-band case only long range correlations $\sim \left|\mathbf{r}-\mathbf{r}'\right|^{-1}$ are present in this temperature region. More detailed analysis of the correlation functions (\ref{eq15}) can be found in Ref. \cite{varg}.

\section{Conclusion}

It was demonstrated that two-gap inhomogeneous superconductivity involves more complicated physics compared to one gap situation. Critical and non-critical coherence lengths, found in two-band scenario, scale the spatial behaviour of superconducting fluctuations as well as their correlation functions. The concrete picture depends substantially on the temperature and the strength of interband interaction. However, even in the vicinity of critical point both short-range and long-range correlations are essential.

\begin{acknowledgements}
This research was supported by the European Union through the European Regional Development Fund (Centre of Excellence "Mesosystems: Theory and
Applications", TK114). We
acknowledge the support by the Estonian Science Foundation, Grant No 8991.
\end{acknowledgements}

\end{document}